\documentstyle[preprint,psfig,tighten,eqsecnum,floats,aps]{revtex}

\newtheorem{Theorem}{Theorem}
\newtheorem{Definition}{Definition}
\newtheorem{Lemma}{Lemma}

\def\Cyl{{\rm Cyl}}
\def\be{\begin{equation}}
\def\ee{\end{equation}}
\def\ba{\begin{eqnarray}}
\def\ea{\end{eqnarray}}
\def\A{{\cal A}}
\def\G{{\cal G}}
\def\ag{{{\cal A}/{\cal G}}}
\def\C{{\cal C}}
\def\o{\overline}
\def\Cb{\o{\C}}
\def\L{{\cal L}}

\def\Comp{{\mathchoice
{\setbox0=\hbox{$\displaystyle\rm C$}\hbox{\hbox to0pt
{\kern0.4\wd0\vrule height0.9\ht0\hss}\box0}}
{\setbox0=\hbox{$\textstyle\rm C$}\hbox{\hbox to0pt
{\kern0.4\wd0\vrule height0.9\ht0\hss}\box0}}
{\setbox0=\hbox{$\scriptstyle\rm C$}\hbox{\hbox to0pt
{\kern0.4\wd0\vrule height0.9\ht0\hss}\box0}}
{\setbox0=\hbox{$\scriptscriptstyle\rm C$}\hbox{\hbox to0pt
{\kern0.4\wd0\vrule height0.9\ht0\hss}\box0}}}}

\def\Co{{\mathchoice
{\setbox0=\hbox{$\displaystyle\rm C$}\hbox{\hbox to0pt
{\kern0.4\wd0\vrule height0.9\ht0\hss}\box0}}
{\setbox0=\hbox{$\textstyle\rm C$}\hbox{\hbox to0pt
{\kern0.4\wd0\vrule height0.9\ht0\hss}\box0}}
{\setbox0=\hbox{$\scriptstyle\rm C$}\hbox{\hbox to0pt
{\kern0.4\wd0\vrule height0.9\ht0\hss}\box0}}
{\setbox0=\hbox{$\scriptscriptstyle\rm C$}\hbox{\hbox to0pt
{\kern0.4\wd0\vrule height0.9\ht0\hss}\box0}}}}

\def\Rl{{\mathchoice 
{\setbox0=\hbox{$\displaystyle\rm R$}\hbox{\hbox to0pt
{\kern0.4\wd0\vrule height0.9\ht0\hss}\box0}}
{\setbox0=\hbox{$\textstyle\rm R$}\hbox{\hbox to0pt
{\kern0.4\wd0\vrule height0.9\ht0\hss}\box0}}
{\setbox0=\hbox{$\scriptstyle\rm R$}\hbox{\hbox to0pt
{\kern0.4\wd0\vrule height0.9\ht0\hss}\box0}}
{\setbox0=\hbox{$\scriptscriptstyle\rm R$}\hbox{\hbox to0pt
{\kern0.4\wd0\vrule height0.9\ht0\hss}\box0}}}}

\def\vp{\varphi}

\def\difs{{\rm Diff}(M,s)}
\def\difo{{\rm Diff}(M,\omega)}

\def\Z{{\bf Z}}

\def\a{\alpha}
\def\rs{\Rl \times S^1}
\def\la{\label}

\def\bi{\begin{itemize}}
\def\ei{\end{itemize}}

\begin{document}

\draft
\title{Constructing Hamiltonian quantum theories from path integrals\\
in a diffeomorphism invariant context}


\author{Abhay\ Ashtekar${}^{1,5}$,  Donald Marolf${}^{2,5}$
Jos\'e Mour\~ao${}^{3}$ and Thomas Thiemann${}^{4,5}$}

\address{1. Center for Gravitational Physics and Geometry \\
Department of Physics, The Pennsylvania State University \\
University Park, PA 16802, USA}

\address{2. Department of Physics, Syracuse University,\\
Syracuse, NY 13244, USA,}

\address{3. Departamento de F\'{\i}sica, Instituto Superior\\
T\'ecnico, Av. Rovisco Pais, 1049-001 Lisboa Codex, Portugal,}

\address{4. Albert-Einstein Institut, MPI f\"ur 
Gravitationsphysik,\\ Am M\"uhlenberg 1, 14476 Golm bei Potsdam, Germany,}

\address{5. Institute for Theoretical Physics,\\ 
University of California, Santa Barbara, California, 93106, USA.}

\maketitle

\begin{abstract}
Osterwalder and Schrader introduced a procedure to obtain a
(Lorentzian) Hamiltonian quantum theory starting from a measure on the
space of (Euclidean) histories of a scalar quantum field. In this
paper, we extend that construction to more general theories which do
not refer to any background, space-time metric (and in which the space
of histories does not admit a natural linear structure). Examples
include certain gauge theories, topological field theories and
relativistic gravitational theories. The treatment is self-contained
in the sense that an a priori knowledge of the Osterwalder-Schrader
theorem is not assumed.

\end{abstract}


\section{Introduction}
\label{s1}

For scalar field theories in flat space-time, the Osterwalder-Schrader
framework provides a valuable link between Euclidean and Minkowskian
descriptions of the quantum field.  In this paper we will focus only on
one aspect of that framework, namely the so-called `reconstruction
theorem' \cite{0} which enables one to recover the Hilbert space of
quantum states and the Hamiltonian operator, starting from an
appropriate measure on the space of Euclidean paths. At least in
simple cases, this procedure provides a precise correspondence between
the path integral and canonical approaches to quantization.  However,
since even the basic axioms of the framework are deeply rooted in
(Euclidean and) Poincar\'e invariance, a priori it is not obvious that
the construction would go through in  diffeomorphism invariant
theories, such as general relativity. In particular, the idea of a
`Wick rotation', implicit in the original framework, has no obvious
meaning in this context.  The purpose of this paper is to show that,
in spite of these difficulties, the construction can be generalized to
such theories.

While diffeomorphism invariance is our primary concern, we will also
address another issue that arises already in Minkowskian field
theories. It stems from the fact that the standard
Osterwalder-Schrader framework \cite{1} is geared to `kinematically
linear' systems ---such as interacting scalar field theories--- where
the space of Euclidean paths has a natural vector space
structure. More precisely, paths are assumed to belong to the space of
Schwartz distributions and this assumption then permeates the entire
framework. Although the assumption seems natural at first, in fact it
imposes a rather severe limitation on physical theories that one can
consider. In particular, in non-Abelian gauge theories, the space of
gauge equivalent connections does {\it not} have a natural vector
space structure. Therefore, if one wishes to adopt a manifestly gauge
invariant approach, the space of histories can not be taken to be one
of the standard spaces of distributions \cite{almmt2,almmt1}. Even if
one were to use a gauge fixing procedure, because of Gribov
ambiguities, one can not arrive at a genuine vector space if the
space-time dimensions greater than two.  Our extension of the
Osterwalder-Schrader reconstruction theorem will incorporate such
`kinematically non-linear' theories.

Let us now return to our primary motivation. As noted above, since the
standard formulation of the reconstruction theorem makes a crucial use
of a flat, background metric, it excludes diffeomorphism invariant
theories. The most notable examples are gravitational theories
such as general relativity and topological ones such as Chern-Simons
and BF theories, in which there is no background metric {\it at
all}. To incorporate these cases, one has to generalize the very
setting that underlies the Osterwalder-Schrader framework.  A natural
strategy would be to substitute the Poincar\'e group used in the
original treatment by the diffeomorphism group. However, one
immediately encounters some technical subtleties. In certain cases,
such as general relativity on spatially compact manifolds, all
diffeomorphisms are analogous to gauge transformations. Hence, while
they have a non-trivial action on the space of paths, they have to act
trivially on the Hilbert space of physical states. In other cases,
such as general relativity in the asymptotically flat context,
diffeomorphisms which are asymptotically the identity correspond to
gauge while those that preserve the asymptotic structure but act
non-trivially on it define genuine symmetries. These symmetries should
therefore lead to non-trivial Hamiltonians on the Hilbert space of
physical states. The desired extension of the Osterwalder-Schrader
framework has to cater to these different situations appropriately.

Thus, from a conceptual viewpoint, the extensions contemplated here
are very significant. However, it turns out that, \textit{once an
appropriate setting is introduced}, the technical steps are actually
rather straightforward. With natural substitutions suggested by this
generalized setting, one can essentially follow the same steps as in
the original reconstruction \cite{1} with minor technical
modifications. In particular, it is possible to cater to the various
subtleties mentioned above.

The plan of this paper is as follows. Section \ref{s2} introduces the
new setting and section \ref{s3} contains the main result, the
generalized reconstruction theorem.  Section \ref{s4} (and the
Appendix) discuss examples which illustrate the reconstruction
procedure.  These examples will in particular suggest a manifestly
gauge invariant approach in the non-Abelian context and also bring out
different roles that the diffeomorphism group can play and subtleties
associated with them. Section \ref{s5} summarizes the main results and
briefly discusses their ramifications as well as limitations.

\section{The general setting}
\label{s2}

This section is divided into three parts. In the first, we introduce
the basic framework, in the second we discuss some subtleties
associated with diffeomorphism invariance and in the third we present
the the modified axioms.

\subsection{Basic framework}
\label{s2.1}

Heuristically, our task is to relate the path integral and canonical
approaches for a system which may not have a background metric
structure.  Let us therefore begin with a differential manifold $M$ of
dimension D+1 and topology $\Rl\times\sigma$, where $\sigma$ is a
D-dimensional smooth manifold of arbitrary but fixed topology. $M$
will serve as the non-dynamical arena for theories of interest.  The
product topology of $M$ will play an important role in what follows.
In particular, it will enable us to generalize the notion of
`time-translations' and `time-reflections' which play an important
role in the construction. Since our goal is to obtain a Hamiltonian
quantum theory, it is not surprising that we have to restrict
ourselves to a product topology.

Our generalized Osterwalder-Schrader axioms will require the use of
several structures associated with $M$.  The fact that $M$ is
diffeomorphic to $\Rl\times\sigma$ in particular means that it can be
foliated by leaves diffeomorphic to $\sigma$. To be precise, consider
the set ${\rm Emb}(\sigma,M)$ of all embeddings of $\sigma$ into $M$.
A {\it foliation} $E=\{E_t\}_{t\in\Rl}$ is a one-parameter family of
elements of ${\rm Emb}(\sigma, M)$, $E_t\in {\rm Emb}(\sigma,M)$,
which varies smoothly with $t$ and provides a diffeomorphism between
$\Rl \times \sigma$ and $M$. The set of foliations ${\rm
Fol}(\sigma,M)$, given by all diffeomorphisms from $\Rl \times \sigma$
to $M$, will be of special interest to us.  Notice that the embedded
hyper-surfaces $\Sigma_t=E_t(\sigma)$ have not been required to be
`space-like, time-like or null.' Indeed, there is no background
metric to give meaning to these labels.

Each foliation $E\in {\rm Fol}(\sigma,M)$ enables one to generalize
the standard notions of time translation and time reflection.  To see
this, first note that since $E$ is of the form $E: \Rl \times \sigma
\rightarrow M; \ (t,x) \mapsto X = E_t(x)$, the inverse map $E^{-1}$
defines functions $t_E(X)$ and $x_E(X)$ from $M$ to $\Rl$ and
$\sigma$ respectively.  The time translation $\varphi^\Delta$, with
$\Delta \in \Rl$, is the diffeomorphism on $M$, $(t_E (X), x_E(X))
\mapsto (t_E(X) + \Delta, x_E(X))$, which is simply a shift of the
time coordinate $t_E$ by $t$, holding $x_E(X)$ fixed.  Similarly, the
time reflection $\theta_E$ is the diffeomorphism of $M$ defined by
$(t_E(X),x_E(X))\mapsto (-t_E(X),x_E(X))$.  We also consider the
positive and negative half spaces $S_E^\pm$, defined by $X \in
S^\pm_E$ if and only if $\pm t_E(X) \ge 0$. Although these notions are
tied to a specific foliation, our final constructions and results will
\textit{not} refer to a preferred foliation.

We now turn to the structures associated with the particular quantum
field theory under consideration.  Let us assume that our theory is
associated with a classical Lagrangian density which depends on a
collection of basic (bosonic) fields $\phi$ on $M$ and their various
partial derivatives.  We will not explicitly display discrete indices
such as tensorial or representation space indices, so that the symbol
$\phi$ may include, in addition to scalar fields, higher spin fields
which may possibly take values in a representation of the Lie-algebra
of a structure group.  The fields $\phi$ belong to a space $\C$ of
{\it classical histories} which is typically a space of smooth
(possibly Lie algebra-valued) tensor fields equipped with an
appropriate Sobolev norm.  In the case of a gauge theory, there will
be an appropriate gauge bundle over $M$.  We assume that the action of
${\rm Diff}(M)$, the diffeomorphism group of $M$, has a lift to this
bundle, from which an action on $\C$ follows naturally.  For
notational simplicity, we will denote this action of ${\rm Diff}(M)$
on $\C$ simply by $\phi\mapsto \varphi \phi$ for any $\varphi \in {\rm
Diff}(M)$.

Of greater interest than $\C$ will be the set ${\Cb}$ of {\it quantum
histories} which is generally an extension of $\C$. In a kinematically
linear field theory, $\Cb$ is typically the space of Schwartz
distributions \cite{1}. The extension from $\C$ to $\Cb$ is essential
because, while $\C$ is densely embedded in $\Cb$ in the natural
topology, in physically interesting cases, $\C$ is generally of
measure zero, whence the genuinely distributional paths in $\Cb$ are
crucial to path integrals. In the more general case now under
consideration, we leave the details of the extension unspecified, as
they depend on the particulars of the theory being considered, and
refer to elements of ${\Cb}$ simply as {\it generalized fields}.  For
example, in a gauge theory these might include generalized connections
discussed briefly in Section \ref{s4} (and in detail in
\cite{almmt2}).  For notational simplicity, the symbol $\phi$ will be
used to denote generalized fields (elements of ${\Cb}$) as
well as smooth fields in ${\C}$; the context will remove the ambiguity.

Consider then a suitable collection of subsets of $\Cb$ and denote by
${\cal B}$ the $\sigma$-algebra it generates.  This equips $\Cb$ with
the structure of a measurable space. Let us further consider the set
${\cal F}(\Cb)$ of measurable functions on this space (that is,
functions for which the pre-image of any Lebesgue measurable set is
a measurable set).

With this background material at hand, we can now introduce a key
technical notion, that of a {\it label set} ${\L}$, which in turn will
enable us to define the basic random variables and stochastic process.
${\L}$ is to be regarded as being `dual' to the space ${\Cb}$ of
generalized fields.  That is, it must be chosen to match the structure
of $\overline{\C}$ so that there is a well-defined `pairing' $P$:
\be \label{1}
P\; :\; {\cal L}\to 
{\cal F}(\overline{{\cal C}});\; f \mapsto P_f\;.
\ee
For example, in kinematically linear field theories on $\Rl^{D+1}$,
typically $\overline{\cal C}$ is taken to be the space of Schwartz
distributions with appropriate tensor and internal indices. Then,
${\L}$ consists of smooth, rapidly decreasing (test) functions $f$ on
$\Rl^{D+1}$, with the pairing $P$ defined by
$P_f(\phi)=\exp(i\phi(f))$, where $\phi(f) :=\int d^{D+1}X\,\,
\phi(X)f(X)$. In $SU(2)$ gauge theories, a natural candidate for
${\L}$ is the space of loops on $M$ and the pairing is then defined by
$P_f(\phi)= {\rm Tr}\, h_f(\phi)$, the trace of the holonomy
$h_f(\phi)$ of the generalized connection $\phi$ around the loop $f$
in a suitable representation of the structure group
\cite{almmt1,almmt2}.  In general, we will assume that each $f \in
{\L}$ is `associated with' a set supp$(f) \subset M$, which we call
the support of $f$. The pairing defines a {\it stochastic process}
$f\to P_f(\phi)$, and we refer to $P_f$ as a {\it random variable}.
\medskip

In the general framework, we will not be concerned with the details
of the pairing $P$, but merely ask that it satisfies  the following three
properties:\\

{\it (A1) The pairing is diffeomorphism covariant, in the sense that
there exists a left action of ${\rm Diff}(M)$ on $\L$, which we 
denote $f\mapsto (\varphi^{-1}) f$, such that $P_f(\varphi \phi)=
P_{(\varphi^{-1}) f}(\phi)$ for any $\phi\in {\C}$.  
Furthermore, we require that {\rm supp}$((\varphi^{-1}) f) 
= \varphi(${\rm supp}$(f))$.}\\
\\

We also introduce a left action of $\varphi$ on random variables:
$\varphi(P_f) = P_{\varphi^{-1} f}$.
Note that in the familiar case of scalar field theories where the
label set is taken to be the set of Schwarz space functions ($f \in {\cal 
S}$),
the action of $\varphi$ on ${\cal L}$ is $(\varphi^{-1}) f 
= f \circ \varphi^{-1}$.

\medskip

Let us denote by ${\cal A}$ the set of finite linear combinations
($N<\infty$) of random variables $P_f:$
\be \label{2}
\psi(\phi):=\sum_{I=1}^N z_I P_{f_I}(\phi)
\ee
with $z_I\in\Co$ and $f_I\in {\L}$.  The
second assumption about the pairing $P$ is :\\
\\
{\it (A2) The vector space $\cal A$ is in fact a $\star$-algebra with unit,
whose $\star$ operation is complex conjugation of functions on $\overline 
{\C}$.  The 
algebraic operations of ${\cal A}$ must
commute with the action of diffeomorphisms in the sense that:
}\\
\begin{equation}
{\it For} \  a, b \in {\cal A}, \ \varphi(ab) 
=[\varphi(a)] [\varphi(b)], \ \ {\it and} \ \  
[\varphi(a^\star)]= [\varphi(a)]^\star.
\end{equation}
The first part of this property will allow us to calculate scalar
products between elements of ${\cal A}$ with respect to suitable
measures on $\overline{{\C}}$ purely in terms of expectation values of
the random variables $P_f$. Note that, in the kinematically linear
theories as well as the gauge theories referred to above, this
assumption is automatically satisfied.

Next, let us consider a $\sigma$-additive probability measure $\mu$ on
the measurable space $(\overline{{\C}},{\cal B})$, thus equipping it
with the structure of a measure space $({\Cb},{\cal B},\mu)$. This
structure naturally gives rise to the so called `history Hilbert
space'
\be \label{3}
{\cal H}_{D+1}:=L_2(\overline{{\C}}, d\mu)
\ee
of square integrable functions. We denote the inner product between
$\psi,\psi'\in {\cal A}$ by 
\be \label{3a}
\langle \psi,\psi' \rangle:=\int_{\Cb}d\mu \ 
\overline{\psi(\phi)}\psi'(\phi).
\ee
Our third requirement on $P$ is that :\\ 
\\ 
{\it (A3) The space $\cal A$ is dense in ${\cal H}_{D+1}$ for some
measure $\mu$ on $\overline{\cal C}$.}\\ 
\\ 
As mentioned above, we are primarily interested in diffeomorphism
invariant theories.  The pairing allows us to define a representation
$\hat{U}(\varphi)$ of ${\rm Diff}(M)$ on the dense subspace ${\cal A}$
of ${\cal H}_{D+1}$:
\be \label{4}
[\hat{U}(\varphi)P_f](\phi):=P_{(\varphi^{-1})f}(\phi)=(\varphi P_f)(\phi).
\ee
At this point, $\hat{U}(\varphi)$ is a densely defined operator on
${\cal H}_{D+1}$. When the measure $d\mu$ is invariant under
diffeomorphisms, we will see that this operator is in fact unitary and
extends to all of ${\cal H}_{D+1}$.

Finally, in the formulation of the key, `reflection positivity'
axiom and in the proof of the reconstruction theorem, we will need
certain subsets ${\cal A}_E^\pm$ of $\cal A$. These are defined by
restricting the supports of the $f_I$ in (\ref{2}) to be contained in
half spaces $S_E^\pm$ on which $\pm t_E \ge 0$.

\medskip

\subsection{Subtleties}
\label{s2.2}

We are nearly ready to formulate our extension of the
Osterwalder-Schrader axioms.  However, our emphasis on diffeomorphism
invariant systems will cause a certain change of perspective from the
familiar case, e.g. of a kinematically linear field theory in flat
space-time.  In these simpler theories, the Hamiltonian is an object
of primary concern, and its construction is central to the
Osterwalder-Schrader reconstruction theorem. Now, we no longer have a
background metric and therefore no a priori notion of time
translations which Hamiltonians normally generate. An obvious strategy
is to treat {\it all} diffeomorphisms as symmetries and seek the
corresponding Hamiltonians on the Hilbert space of physical
states. However, this turns out not to be the correct procedure
because of two subtleties.

First, in many diffeomorphism invariant systems, the structure of the
classical theory tells us that all diffeomorphisms should be regarded
as gauge transformations. This can follow from one of the following
three related considerations: i) The initial value formulation could
show that the initial data can determine a classical solution {\it
only} up to diffeomorphisms; or, ii) For fixed boundary values of
fields, the variational principle may provide infinitely many
solutions, all related to one another by diffeomorphisms which are
identity on the boundary; or, iii) In the Hamiltonian formulation,
there may be first class constraints whose Hamiltonian flows
correspond to the induced action of ${\rm Diff}(M)$ on the phase
space.  (Typically, one of these implies the other two.)  An example
where this is the case is general relativity on a spatially compact
manifold. In these cases, one expects quantum states to be
diffeomorphism invariant, i.e., the corresponding Hamiltonian
operators to vanish identically on the physical Hilbert space. In
these theories, then, the reconstruction problem should reduce only to
the construction of the Hilbert space of physical states starting from
a suitable measure on $\Cb$.

The second subtlety is that many interesting theories use a background
structure. They are therefore {\it not} invariant under the full
diffeomorphism group but only under the sub-group which preserves the
background structure.  An interesting example is provided by the
Yang-Mills theory in two dimensional space-times which requires an
area form, but not a full metric, for its formulation.  The theory is
therefore invariant under the group of all area preserving
diffeomorphisms, a group which is significantly larger than, say, the
Poincar\'e group but smaller than the group of {\it all}
diffeomorphisms. (Since the area form is a symplectic form in two
dimensions, the group of area preserving diffeomorphisms coincides
with the group of symplectomorphisms.)  A more common situation is
illustrated by general relativity in any space-time dimension, subject
to asymptotically flat (or anti-de Sitter) boundary conditions. Here,
the background structure consists of a flat (or anti-de Sitter)
geometry at infinity. One must therefore restrict oneself to those
diffeomorphisms which preserve the specified asymptotic structure.  In
the general case, we will denote the background structure by $s$ and
the sub-group of ${\rm Diff}(M)$ preserving this structure by%
\footnote{If there is no background structure, as for example in
general relativity or topological field theories on a spatially
compact manifold, by ${\rm Diff}(M,s)$ we will mean simply ${\rm
Diff}(M)$.}
${\rm Diff}(M, s)$.  

In the presence of a background structure $s$, our foliations will
also be restricted to be compatible with $s$ in the sense that the
associated generalized time translations $\varphi_E^t$ and time
reflections $\theta_E$ constructed above preserve $s$.  Now, given any
two foliations $E,\tilde{E}$, there is a unique diffeomorphism
$\varphi_{E\tilde{E}}$ on $M$ which maps $E$ to $\tilde{E}$. Note
however that even if $E$ and $\tilde{E}$ are compatible with $s$,
$\varphi_{E,\tilde{E}}$ need not preserve $s$. This leads us to the
following important definitions:
\begin{Definition} \label{def1}
(a) Two foliations $E$ and $\tilde{E}$ are {\it strongly 
equivalent} if $\tilde{E}= \varphi_{E \tilde{E}}\circ E$ for some
$\varphi_{E \tilde{E}} \in {\rm Diff}(M,s)$.\\
(b) Two foliations $E, \tilde{E}$ will be said to
be {\it weakly} equivalent if there exists foliations $E', \tilde{E}'$
which are strongly equivalent to $E, \tilde{E}$ respectively such that
the time-reflection maps of $E'$ and $\tilde{E}'$ coincide, i.e.
$\theta_{E'}= \theta_{\tilde{E}'}$.
\end{Definition}

Note that strong equivalence trivially implies weak equivalence but
the converse is not true.  A simple example which illustrates the
difference between these two notions of equivalence is provided by
setting $M=\Rl^4$ and choosing the background structure $s$ to be a
Minkowskian metric $\eta$.  Define $E, \tilde{E}$ as follows: $E:
\Rl\times \Rl^3 \rightarrow M; \,(t, x) \mapsto E_t(x) = (t, x)$ and
$\tilde{E}: \Rl\times \Rl^3 \rightarrow M; \,(t, x) \mapsto
\tilde{E}_t(x) = (bt, x)$, where $b$ is a positive constant. Both of
these foliations are compatible with the background
structure. However, since the diffeomorphism $\varphi_{E \tilde{E}}$
is not an isometry of $\eta$, the two foliations are {\it not}
strongly equivalent. However, they define the same time-reflection map
and are therefore weakly equivalent.

Strong equivalence of $E$ and $\tilde E$ means that the foliations are
in fact related by a symmetry $\varphi_{E \tilde E}$ of the theory and
we will see that this symmetry defines a unitary mapping of the
physical Hilbert space associated with $E$ to that associated with
$\tilde E$ which takes the Hamiltonian generator of $\varphi_E^t$ to
that of $\varphi_{\tilde E}^t$.  In the case of weak equivalence, the
foliations are not related by a symmetry and we should expect no
correspondence between the Hamiltonians.  The point of this
definition, however, is that the construction of the physical Hilbert
space itself will depend only on the time inversion map $\theta_E$
induced by the foliation $E$.  Thus, when $E$ and $\tilde E$ are
weakly equivalent, we will still be able to show that the physical
Hilbert spaces are naturally unitarily equivalent, though this
equivalence will not of course map the generator of $\varphi^t_E$ to
that of $\varphi^t_{\tilde E}$.

Finally, it is typical in such theories that certain diffeomorphisms
play the role of genuine symmetries while others play the role of
gauge.  Generally, there is a normal subgroup ${\rm Diff}_G(M,s)$ of
${\rm Diff}(M,s)$ which acts as gauge while the quotient, ${\rm
Diff}(M,s)/{\rm Diff}_G(M,s)$, acts as a symmetry group.  (In
asymptotically flat general relativity, for example, ${\rm
Diff}_G(M,s)$ consists of asymptotically trivial diffeomorphisms and
the quotient is isomorphic to the Poincar\'e group.)  In these
contexts, we have a mixed situation: ${\rm Diff}_G(M,s)$ should have a
trivial action on the physical Hilbert space, while the action of a
symmetry diffeomorphism should be generated by a genuine Hamiltonian
as in the original reconstruction theorem.

\subsection{Generalized Osterwalder-Schrader Axioms}
\label{s2.3}

With these subtleties in mind, we can now state our generalization of
the Osterwalder-Schrader axioms. Our numbering of the axioms below is
chosen to match that of \cite{1}. For flexibility, we wish to allow
the possibility that a quantum theory may satisfy only a subset of the
following axioms.  We will be careful in what follows to explicitly
state which axioms are required in order that the various conclusions
hold.
 
As in the original construction, the key mathematical object will be a
measure $\mu$ on the space of quantum histories. In Minkowskian field
theories, $\mu$ can be thought of as a rigorous version of the
heuristic measure $\exp -S(\phi)\, {\cal D}\phi$ constructed from the
Euclidean action. In the standard Osterwalder-Schrader framework,
there are two axioms which are central to the construction of the
Hilbert space of states and both are restrictions on the measure
$\mu$.  The first asks that $\mu$ be Euclidean invariant and the
second asks that it satisfy a technical condition called `reflection
positivity' formulated in terms of the `time-reflection' operator
$\theta$ in the Euclidean space. Given a measure $\mu$ with these
properties, one can quotient the space $L^2(\Cb, d\mu)$ of
square-integrable functions on quantum histories by a certain
sub-space, defined by $\theta$, to obtain the Hilbert space of quantum
states.  Heuristically, the restrictions of quantum histories to the
D-dimensional, $t=0$ slice in the Euclidean space define the `quantum
configuration space' ${\Cb}_{t=0}$ and the quotient enables one to
pass from $L^2(\Cb, d\mu)$ to the space of square-integrable functions
on ${\Cb}_{t=0}$. The remaining axioms ensure the existence of a
Hamiltonian operator and existence and uniqueness of the vacuum state.

In the present context, the time reflection operator $\theta$ is
replaced by its generalization $\theta_E$ associated with a foliation
and the Poincar\'e group is replaced by ${\rm Diff}(M,s)$. Thus, given
any foliation $E$, one can essentially repeat the original
construction to obtain a Hilbert space ${\cal H}_D^E$ of physical
quantum states. The diffeomorphism invariance of $\mu$ then provides
unitary maps relating physical Hilbert spaces constructed from equivalent 
foliations.

\begin{Definition} \label{d1} 
A quantum theory of fields $\phi\in\Cb$ on a space-time $M$
diffeomorphic to $\Rl \times\sigma$ is defined by a probability
measure $\mu$ on $\Cb$ and a pairing $P$ satisfying (A1), (A2)
and (A3) above.  The generating functional $\chi$ defined by
\be \label{5}
\chi(f):=<P_f>:=\int_{\Cb} d\mu(\phi) P_f(\phi).
\ee
should satisfy at least the first two of the following axioms:
\begin{itemize}
\item[(II)] DIFFEOMORPHISM INVARIANCE\\
The measure is diffeomorphism invariant%
\footnote{ In the case when there are symmetries in the quantum field 
theory other than diffeomorphism invariance, it would be natural to
ask that the measure be invariant under these symmetries as well.} 
in the sense that, for any $\varphi\in {\rm Diff}(M,s)$, $\chi$
satisfies:
\be \chi(f)=\chi(\varphi^{-1}  f). \ee
\item[(III)] REFLECTION POSITIVITY\\
Consider the sesquilinear form on ${\cal A}_E^+$ defined for any
$E\in {\rm Fol}(\sigma,(M,s))$ by
\be \label{6}
(\psi,\psi')_E:=\langle \hat{U}(\theta_E)\psi,\psi'\rangle \;.
\ee
We require $(\psi,\psi)_E\ge 0$ for any $\psi\in {\cal A}_{E}^+$.
\item[(GI)]GAUGE INVARIANCE\\
For all $\varphi\in {\rm Diff}_G(M,s)$ and all $\psi\in 
{\cal A}_{E}^+$ we require
\footnote{Again, if there exist gauge symmetries in addition to gauge
diffeomorphisms, we ask that these be represented trivially
on the physical Hilbert space as well.}
\be ||\left( \hat{U}(\varphi) - \hat{1}\right) \psi||_E =0,\ee
where $|| \ ||_E$ denotes the norm associated with the inner product
introduced in axiom III.
\item[(I)] CONTINUITY\\
For any $E\in {\rm Fol}(\sigma,(M,s))$ for which the one-parameter
group of diffeomorphisms $\varphi_E^t$ does not belong to ${\rm
Diff}_G(M,s)$, it acts strongly continuously by operators
$\hat{U}(\varphi^t_E)$ on ${\cal H}_{D+1}$.
\item[(IV)] CLUSTERING\\
For any $E\in {\rm Fol}(\sigma,(M,s))$ for which $\varphi_E^t$ does not
belong to ${\rm Diff}_G(M,s)$, we have that
\be \label{7}
\lim_{t\to\infty} \langle \psi,\hat{U}(\varphi^t_E)\psi'\rangle
=\langle \psi,1 \rangle \langle 1,\psi' \rangle
\ee
for any two $\psi,\psi'\in {\cal A}$.
\end{itemize}
\end{Definition}

Note that if axiom (III), (GI), (I) or (IV) holds for some foliation
$E\in {\rm Fol}(\sigma, (M,s))$ then, because of the diffeomorphism
invariance (II) of the measure, the axiom also holds for any $\tilde
E$ which is strongly equivalent to $E$.  Furthermore, if axiom (III)
or (GI) holds for some foliation $E\in {\rm Fol}(\sigma, (M,s))$ then
it in fact holds for any $\tilde E$ which is {\it weakly} equivalent
to $E$.

We will conclude this sub-section by comparing these axioms with the
standard ones of Osterwalder-Schrader \cite{1}.  In axiom (II), we
have replaced the Euclidean group in the standard formulation by ${\rm
Diff}(M,s)$, and in (I), the time translation group by $\varphi^t_E$.
Axiom (I) is usually referred to as the `regularity' axiom and
typically phrased as a more technical condition specific to scalar
field theories on a flat background \cite{1}. However, as its
essential role in that case is to ensure strong continuity of time
translations, and as this condition is straightforward to state in the
diffeomorphism invariant context, we have chosen to promote this
condition itself to the axiom.  Finally, note that we have discarded
the zeroth `analyticity axiom' of \cite{1} which, roughly speaking,
requires the generating functional $\chi$ to be analytic in $f\in
{\cal L}$.  However, it is not clear that a label space $\cal L$
appropriate to a diffeomorphism invariant or kinematically non-linear
field theory should carry an analytic structure.  In the standard
formulation this axiom allows one to define Schwinger n-point
functions in terms of $\chi(f)$.  Fortunately, Schwinger functions are
not essential to our limited goal of defining the Hilbert space
theory.

\section{Recovery of the Hilbert Space Theory}
\label{s3}

The central subject of this section is the following straightforward 
extension of the classical Osterwalder-Schrader reconstruction theorem 
\cite{1}.
\begin{Theorem} \label{th1}
i) For each $E\in {\rm Fol}(\sigma,(M,s))$, axioms (II) and (III)
imply the existence of a Hilbert space ${\cal H}^E_D$ of physical
states.  There is a natural class of unitary equivalences between
${\cal H}^E_D$ and ${\cal H}^{\tilde{E}}_D$ for all $E$ and
$\tilde{E}$ in the (weak) equivalence class of $E$.

ii) Axioms (I), (II) and (III) imply the existence of self-adjoint
operators $[\hat{H}^E]$ on ${\cal H}^E_D$ which generate time
translations and which have $[1]_E$ as a vacuum state%
\footnote{{\rm That is, a state annihilated by $[\hat{H}^E]$.
The brackets on $[\hat{H}^E]$ denote an operator on ${\cal H}^E_D$ as
opposed to $L^2({\Cb}, d\mu)$ and $[1]_E \in {\cal H}^E_D$ is the
equivalence class of elements of ${\cal H}^E_{D+1}$ to which the unit
function belongs. See the observation below Eq. (\ref{15}).}}.
If the foliations $E$ and $\tilde{E}$ are strongly equivalent, then
the operators $[\hat{H}^E]$ and $[\hat{H}^{\tilde{E}}]$ are mapped to
each other by a unitary equivalence of the Hilbert spaces ${\cal
H}^E_D$ and ${\cal H}^{\tilde{E}}_D$

iii) Axioms (I), (II), (III) and (IV) imply that the vacuum
vacuum $[1]_E$ is unique in each ${\cal H}^E_D$. These states are
mapped to each other by the unitary equivalence of ii) above.

\end{Theorem}
We break the proof of this theorem into several lemmas. As noted in
the Introduction, the essence of the proof is the same as that in
the original Osterwalder-Schrader reconstruction but we present it
here for completeness. In the following, $E$ is an arbitrary but
fixed foliation compatible with the background structure (if any).
\begin{Lemma} \label{la1}
By axiom (II), the family of operators, densely defined on
${\cal H}_{D+1}$ for any $\varphi\in {\rm Diff}(M,s)$ by 
\be \label{8}
\hat{U}(\varphi)P_f:=P_{(\varphi^{-1}) f}
\ee
can be extended to a unitary representation of ${\rm Diff}(M,s)$. 
\end{Lemma}
Proof of lemma \ref{la1} :\\
By (A3), $\cal A$ is in fact dense in ${\cal H}_{D+1}$. 
Since $\hat{U}(\varphi)$ has the inverse $\hat{U}(\varphi^{-1})$
on $\cal A$, it will be
sufficient to show that $\hat{U}(\varphi)$ is norm-preserving on $\cal A$
for any $\varphi\in {\rm Diff}(M,s)$ and then to use continuity to
uniquely extend it to ${\cal H}_{D+1}$.  Recalling condition (A2), 
for states $a, b \in {\cal A}$, we have 
\begin{equation}
\langle \hat{U}(\varphi) a , \hat{U}(\varphi) b \rangle =
\int_{\Cb}\, d\mu\, \varphi(a^\star b) \, . 
\end{equation}
But, since the measure is diffeomorphism invariant, this is just
the expectation value of $a^\star b$, which is the inner product 
$\langle a, b \rangle$.

$\Box$\\
\begin{Lemma} \label{la2}
By axioms (II), (III) the sesquilinear form (\ref{6}) defines a
non-negative hermitian form on ${\cal A}^+$.
\end{Lemma}
Proof of lemma \ref{la2} :\\
The hermiticity follows easily from the fact that
$\theta_E\circ\theta_E=\mbox{id}_M$ (so that
$[\hat{U}(\theta_E)]^\dagger = \hat{U}(\theta_E)$), unitarity of the
$\hat{U}(\varphi)$ as established in lemma \ref{la1}, and the
hermiticity of $\langle.,.\rangle$. We have \ba \label{11}
\overline{(\psi,\psi')_E} &=& \overline{\langle
\hat{U}(\theta_E)\psi,\psi' \rangle} \ = \ \langle
\psi',\hat{U}(\theta_E)\psi \rangle \nonumber\\ &=& \langle
\hat{U}(\theta_E)\psi',\psi \rangle \ = \ (\psi',\psi)_E \ea
Non-negativity is the content of axiom (III).\\ $\Box$\\
\begin{Lemma} \label{la3}
\label{la4}
The null space ${\cal N}_E:=\{\psi\in {\cal A}_E^+;\;
(\psi,\psi)_E=0\}$ is in fact a linear subspace of ${\cal A}_E^+$
owing to axioms (II), (III).  As a result, 
the form $(.,.)_E$ is well-defined and positive definite on
\be \label{12}
{\cal H}^E_D:=\overline{{\cal A}_E^+/{\cal N}_E},
\ee
where the over-line denotes completion with respect to $(.,.)_E$.
\end{Lemma}
Proof of lemma \ref{la3} :\\
This is a consequence of the Schwarz inequality for positive 
semi-definite, hermitian, sesquilinear forms.\\
$\Box$\\

\begin{Lemma} \label{la5}
The map $[.]_E:\; {\cal A}_E^+ \to {\cal A}_E^+/{\cal N}_E$ is a
contraction,
that is, $||[\psi]_E||_{{\cal H}^E_D}\le ||\psi||_{{\cal H}_{D+1}}$
owing to axioms (II), (III).
\end{Lemma}
Proof of lemma \ref{la5} :\\
By the Schwarz-inequality for $\langle.,.\rangle$ and the unitarity of
$\hat{U}(\theta_E)$, we have \ba \label{15} (||[\psi]||_{{\cal
H}^E_D})^2 &=& |\langle \hat{U}(\theta_E)\psi,\psi \rangle|^2
\nonumber\\ &\le& ||\hat{U}(\theta_E)\psi||_{{\cal H}_{D+1}}
||\psi||_{{\cal H}_{D+1}} = ||\psi||^2_{{\cal H}_{D+1}}.  \ea
$\Box$

We will also need the following observation:
consider an operator $\hat{A}$ on ${\cal H}_{D+1}$ with dense domain
${\cal D}(\hat{A})$ satisfying the following properties.  :\\
Pi) ${\cal D}_E(\hat{A}):=({\cal D}(\hat{A})\cap{\cal A}_E^+)/{\cal N}_E$
is dense in ${\cal H}^E_D$,\\
Pii) $\hat{A}$ maps ${\cal D}(\hat{A})\cap{\cal A}_E^+$ into
${\cal A}_E^+$ and\\
Piii) $\hat{A}$ maps ${\cal D}(\hat{A})\cap{\cal N}_E$ into
${\cal N}_E$.\\
Then the operator $[\hat{A}]:\;{\cal D}_E(\hat{A})\to {\cal H}_D^E$ 
defined by $[\hat{A}][\psi]:=[\hat{A}\psi]$ is well-defined since
$[\hat{A}\psi]=[0]$ for any $\psi\in {\cal N}_E$.

We have now prepared all the tools necessary to complete the proof of
the theorem. Notice that lemmas \ref{la1} through \ref{la5} have so
far used only the axioms (II) and (III).
\\ \\ 
Proof of theorem \ref{th1} :\\
i) For every $E\in {\rm Fol}(\sigma, (M,s))$ we have already
constructed ${\cal H}_D^E$. Now, let us suppose two foliations $E$ and
$\tilde{E}$ are weakly equivalent. Then, there exist $\varphi_{EE'},
\varphi_{\tilde{E} \tilde{E}'} \in {\rm Diff}(M,s)$ such that $E' =
\varphi_{EE'}\circ E$ and 
$\tilde{E}' = \varphi_{\tilde{E}\tilde{E}'}\circ \tilde{E}$ define the same 
time-reflection map. Therefore, ${\cal
H}_D^{E'} = {\cal H}_D^{\tilde{E}'}$. On the other hand, because of
the diffeomorphism $\varphi_{EE'}$, any vector $\psi'\in {\cal
A}^+_{E'}$ is of the form $\hat{U}(\varphi_{EE'})\psi$ for some vector
$\psi\in {\cal A}_E^+$. Since $\hat{U}(\varphi_{EE'})$ maps ${\cal
N}_E$ to ${\cal N}_{E'}$, $\hat{U}(\varphi_{EE'})$ respects the
quotient construction and we can define a norm-preserving operator
$[\hat{U}_{EE'}]:\;{\cal H}^E_D\to {\cal H}^{E'}_D$ by
\be \label{16}
[\hat{U}_{EE'}][\psi]_E:=[\hat{U}(\varphi_{EE'})\psi]_{E'}.
\ee
Thus, the Hilbert spaces ${\cal H}^E_D$, ${\cal H}^{E'}_D$ are
unitarily equivalent in a natural way. Similarly, there exists 
a unitary map $[\hat{U}_{\tilde{E} \tilde{E}'}]: {\cal H}_D^{\tilde{E}}
\rightarrow {\cal H}_D^{\tilde{E}'}$. Hence,  
\be [\hat{U}_{E \tilde{E}}]:= [\hat{U}_{\tilde{E}\tilde{E}'}]^{-1}\,
[\hat{U}_{EE'}] \ee 
is a natural isomorphism between ${\cal H}_D^E$ and ${\cal
H}_D^{\tilde{E}}$.
\medskip

ii)
We will first show the following :
\begin{Lemma} \label{la6}
For any $E\in {\rm Fol}(\sigma,(M,s))$ and any $t\ge 0$ the operator
$\hat{U}(\varphi_E^t)$ satisfies the properties Pi), Pii) and Piii)
above and gives rise to a self-adjoint,
one-parameter contraction semi-group $[\hat{C}^t_E]$ on ${\cal
H}^E_D$. If $\varphi_E^t\not\in\mbox{Diff}_G(M,s)$ then the contraction 
semi-group is also strongly continuous and its generator $[\hat{H}^E]$ is a 
self-adjoint, positive
semi-definite operator on ${\cal H}^E_D$ with $[1]_E$ a vacuum state
for $[\hat{H}^E]$.
\end{Lemma}

Proof of lemma \ref{la6} :\\
First we show that $\hat{U}(\varphi^t_E)$ has the required properties 
Pi), Pii), Piii). Consider $f\in {\cal L}$ with supp$(f)\subset S_E^+$.
Then supp$((\varphi^t_E)^{-1} f)=\varphi^t_E(\mbox{supp}(f))
 \subset \varphi^t_E(S_E^+)=S_E^+$ since for positive $t$ we have a time
translation into $S_E^+$ (`the future of $E_0(\sigma)$').
Thus, $\hat{U}(\varphi^t_E)$ maps ${\cal A}_E^+$ into itself. From
the Schwarz inequality we infer that it also maps ${\cal N}_E$ into itself
and, finally, since ${\cal A}_E^+\subset{\cal A}
={\cal D}(\hat{U}(\varphi^t_E))$ we have 
${\cal D}_E(\hat{U}(\varphi^t_E))={\cal A}_E^+/{\cal N}_E$ which is
dense in ${\cal H}^E_D$. Thus, the operator
\be \label{17}
[\hat{C}^t_E][\psi]_E:=[\hat{U}(\varphi^t_E)\psi]_E 
\ee
is well defined on a dense domain ${\cal D}_E([\hat{C}^t_E]):={\cal
A}_E^+/{\cal N}_E$ of ${\cal H}^E_D$ independent of $t$.

That it defines a semi-group follows from the definition (\ref{17}),
the fact that the $\varphi^t_E$ form a group under composition and the
fact that $\hat{U}$ defines a unitary representation of ${\rm
Diff}(M,s)$ on ${\cal H}_{D+1}$. We have
\be \label{18}
[\hat{C}^t_E][\hat{C}^s_E][\psi]_E=
[\hat{U}(\varphi^t_E)\hat{U}(\varphi^s_E)\psi]_E
=[\hat{U}(\varphi^t_E\circ\varphi^s_E)\psi]_E
=[\hat{U}(\varphi^{t+s}_E)\psi]_E
=[\hat{C}^{t+s}_E][\psi]_E\;.
\ee
Next we show that the $[\hat{C}^t_E]$ are Hermitian on ${\cal H}_D^E$.
For any $\psi,\psi'\in {\cal D}_E([\hat{C}^t_E])$ we have
\ba \label{19}
& & ([\hat{C}^t_E][\psi]_E,[\psi']_E)_E =
([\hat{U}(\varphi^t_E)\psi]_E,[\psi']_E)_E=
\langle \hat{U}(\theta_E)\hat{U}(\varphi^t_E)\psi,\psi' \rangle
\nonumber\\
& = & \langle \hat{U}(\varphi^t_E)^\dagger\hat{U}(\theta_E)\psi,\psi'
\rangle =\langle \hat{U}(\theta_E)\psi,\hat{U}(\varphi^t_E)\psi'
\rangle =([\psi]_E,[\hat{C}^t_E][\psi']_E)_E \ea
where we have used
$\theta_E\circ\varphi^t_E=\varphi^{-t}_E\circ\theta_E$ and the
unitarity of the representation of the diffeomorphism group on ${\cal
H}_{D+1}$.  In particular, we see that ${\cal D}_E([\hat{C}^t_E])$ is
contained in ${\cal D}_E([\hat{C}^t_E]^\dagger)$.

The contraction property now follows from hermiticity:
For $\psi\in {\cal A}_E^+$, we use reflection positivity and find that
\ba \label{20}
0 &\le& ||[\hat{C}^t_E][\psi]_E||_{{\cal H}^E_D}=
([\psi]_E,[\hat{C}^{2t}_E]
[\psi]_E)_E^{1/2}
\nonumber\\
&
\le& ||[\psi]_E||_{{\cal H}^E_D}^{1/2}
||[\hat{C}^{2t}_E][\psi]_E||_{{\cal H}^E_D}^{1/2}
\ea
from the Schwarz inequality. We see, in particular from the first line
of (\ref{20}) that $[\hat{C}^t_E]$ is a positive semi-definite
operator. Iterating $n$-times we arrive at
\be \label{21}
||[\hat{C}^t_E][\psi]_E||_{{\cal H}^E_D}
\le 
||[\psi]_E||_{{\cal H}^E_D}^{\sum_{k=1}^n(1/2)^k}
||[\hat{C}^{2^nt}_E][\psi]_E||_{{\cal H}^E_D}^{(1/2)^n}\;.
\ee
Now, using lemma \ref{la5} and again the unitarity of the representation
of the diffeomorphism group on ${\cal H}_{D+1}$ we have
$||[\hat{C}^{2^nt}_E][\psi]_E||_{{\cal H}^E_D}\le
||\psi||_{{\cal H}_{D+1}}$
and finally since $\sum_{k=1}^n(1/2)^k=1-(1/2)^n$, we find
\be \label{22}
||[\hat{C}^t_E][\psi]_E||_{{\cal H}^E_D}
\le 
||[\psi]_E||_{{\cal H}^E_D}^{1-(1/2)^n}
||\psi||_{{\cal H}_{D+1}}^{(1/2)^n}\;.
\ee
Taking the limit $n\to\infty$ of (\ref{22}) we find the desired result
\be \label{22a}
||[\hat{C}^t_E][\psi]_E||_{{\cal H}^E_D}
\le ||[\psi]_E||_{{\cal H}^E_D}\;.
\ee
The hermiticity of $\hat{C}^t_E$ together with its boundedness implies
that it can be extended to all of ${\cal H}^E_D$ as a self-adjoint,
positive semi-definite operator.

So far we have made no use of axiom (I). However, it is this axiom
that guarantees the existence of a generator for $[\hat C^t_E]$.  Note
that, for any $\psi\in {\cal H}^E_D$, using the hermiticity of
$[\hat{C}^t_E]$, we have
\ba \label{23}
0&\le& ||[\hat{C}^t_E][\psi]_E-[\psi]_E||_{{\cal H}^E_D}^2
\nonumber\\
&=&([\psi]_E,[\hat{C}^{2t}_E][\psi]_E)_E
+([\psi]_E,[\psi]_E)_E
-2([\psi]_E,[\hat{C}^t_E][\psi]_E)_E
\nonumber\\
&=&|\langle \hat{U}(\theta_E)\psi,(\hat{U}(\varphi^{2t}_E)+1-
2\hat{U}(\varphi^t_E))\psi \rangle|
\nonumber\\
&\le&| \langle \hat{U}(\theta_E)\psi,(\hat{U}(\varphi^{2t}_E)-1)\psi \rangle|
+2| \langle \hat{U}(\theta_E)\psi,(\hat{U}(\varphi^t_E)-1)\psi \rangle|
\nonumber\\
&\le& ||\psi||_{{\cal H}_{D+1}}
[||(\hat{U}(\varphi^{2t}_E)-1)\psi||_{{\cal H}_{D+1}}
+2||\hat{U}(\varphi^t_E)-1)\psi||_{{\cal H}_{D+1}}]\;.
\ea
Using axiom (I), strong continuity of the one parameter group of
unitary operators $\hat{U}(\varphi_E^t)$, we find that the limit of
(\ref{23}) vanishes as $t\to 0$, establishing strong continuity of the
one-parameter self-adjoint contraction semi-group on ${\cal H}^E_D$.
Therefore, using the Hille-Yosida theorem \cite{6} we infer that
$[\hat{C}^t_E]=\exp(-t[\hat{H}^E])$ where the generator $[\hat H^E]$
is a positive semi-definite operator 
on ${\cal H}^E_D$. It must annihilate the
state $[1]_E$ as $[\hat{C}^t_E][1]_E=[\hat{U}(\varphi^t_E)1]_E=[1]_E$
for any $t\ge 0$.\\ $\Box$\\
\\

Clearly, if foliations $E,\tilde{E}\in {\rm Fol}(\sigma,(M,s))$
are strongly equivalent, their generators are related by
\be \label{24}
[\hat{H}^{\tilde{E}}]=[\hat{U}(\varphi_{E\tilde{E}})][\hat{H}^E]
[\hat{U}(\varphi_{E\tilde{E}}^{-1})].
\ee
\\
iii)\\ 
So far, axiom (IV) has not been invoked. Axiom (IV) tells us that the
limit \,\, $\lim_{t\to\infty}\hat{U}(\varphi^t_E)$ becomes the
projector $|1 \rangle \langle  1|$ in the weak operator topology. Suppose that
there exists a state $\Omega_E$ which is orthogonal to $|1\rangle$ and
satisfies $\hat{U}(\varphi^t_E)\Omega_E=\Omega_E$ for any
$t\ge 0, E\in {\rm Fol}(\sigma,(M,s))$. We then have
\be \label{25}
||\Omega_E||^2_{{\cal H}_{D+1}}\, = \,
\lim_{t\to\infty} \langle \Omega_E,\hat{U}(\varphi^t_E)\Omega_E \rangle
=| \langle 1,\Omega_E \rangle|^2=0.
\ee
This demonstrates the uniqueness of the vacuum and
concludes the proof of the theorem.\\
$\Box$
\\

We will conclude this section with a few remarks.

a) The uniqueness result in part ii) of the theorem can be slightly
extended. Let $E$ and $\tilde{E}$ be weakly equivalent (rather than
strongly, as required in part ii)). Then, there exist $\varphi_{EE'},
\varphi_{\tilde{E}\tilde{E}'} \in {\rm Diff}(M,s)$ such that
$\theta_{E'} = \theta_{\tilde{E}'}$. Although, in general the
diffeomorphism $\varphi_{EE'}\circ \varphi^{-1}_{\tilde{E}\tilde{E'}}$
will not map the foliation $E$ to $\tilde{E}$, it may map the time
translation $\varphi_E^t$ to the time translation
$\varphi_{\tilde{E}}^t$. In this case, the unitary map
$U_{E,\tilde{E}}$ of part i) of the Theorem will map $[\hat{H}^E]$ to
$[\hat{H}^{\tilde{E}}]$ as in (\ref{24}).

b) At first it may seem surprising that the uniqueness result for the
Hamiltonian is not as strong as in the standard Osterwalder-Schrader
construction. However, this is to be expected on general grounds. In
the standard construction, the notion of time translation is rigid. In
the present context, there is much more freedom. As shown by the
example at the end of section \ref{s2.2}, from the viewpoint of the
general framework, already in Minkowski space we are led to allow both
$\partial / \partial t$ and $b\, \partial/\partial t$ as time
evolution vector fields for any positive constant $b$. Clearly, the
Hamiltonian operators must also differ by a multiplicative constant in
this case. More generally, agreement between Hamiltonians can be
expected only if the two generate the same (or equivalent)
`translations' in space-time. This is precisely what our uniqueness
result guarantees.

c) If two foliations $E,\tilde{E}$ are not even weakly equivalent,
there may be no natural isomorphism between the physical Hilbert
spaces ${\cal H}_D^E$ and ${\cal H}_D^{\tilde{E}}$. In the specific
examples we will consider in the next section, all foliations will in
fact be weakly equivalent. However, due to the so-called
`super-translation freedom' \cite{st}, the situation may well be
different in asymptotically flat general relativity in four space-time
dimensions.  It would be interesting to find explicit examples in
which this inequivalence occurs and to understand its physical
significance.

d) Note that, as in the original Osterwalder-Schrader reconstruction
theorem, the Hilbert space theory obtained here is not as complete as
one would ideally like it be. In particular, no prescription has been
given to construct quantum operators corresponding to the classical
(Dirac) observables.

\section{Examples}
\label{s4}

In this section we discuss three examples of measures on spaces of
quantum histories.  The first example is natural from a mathematical
viewpoint but does not obviously come from the path integral
formulation of a theory of direct physical interest.  (However, we
show in the Appendix that, if $D=1$, this measure is naturally
associated with a universality class of generally covariant quantum
gauge field theories. See \cite{12} for further details.) This
measure satisfies some of our axioms. The other two measures satisfy
all of our axioms and come from the following systems: Yang-Mills
theory in two space-time dimensions and general relativity in three
space-time dimensions (or, B-F theory in any space-time dimension). In
all three cases, the space $\Cb$ of quantum histories is kinematically
non-linear and there is no background metric. These examples serve to
bring out different aspects of the generalization of the
reconstruction theorem.

\subsection{The Uniform Measure for Gauge Theories}
\label{s4.1}

The space of `generalized connections' admits a natural
diffeomorphism invariant measure in any space-time dimension, which we
will refer to as the {\it uniform measure} and denote by $\mu_{0}$
\cite{7}. It plays a crucial role in the kinematical part of a
non-perturbative approach to the quantization of diffeomorphism
invariant theories of connections.%
\footnote{See \cite{almmt3} for a summary and a discussion of the
mathematical details. This approach was motivated in large measure by
ideas introduced in \cite{GT} and \cite{CL}.}
It has also led to a rich quantum theory of geometry
\cite{qg}.

As remarked above, if $D\not=1$, it is unlikely that $\mu_0$ would
arise as the measure on the space of quantum histories in a theory of
direct physical interest. Nonetheless, we discuss it here as a
simple example of a measure which satisfies axioms II,III and IV above
and to illustrate the construction of the Hilbert space ${\cal
H}_D^E$. (For a more complete discussion of this measure, see
\cite{almmt3,7}.)

Suppose we are interested in a theory of connections based on a
compact structure group $K$ on a D+1 dimensional space-time manifold.
For simplicity, in this brief account we will set $K=SU(2)$, assume
that the principal $K$-bundle over $M$ is trivial, and work with a
fixed trivialization. There is no background structure and so ${\rm
Diff}(M,s)$ is just ${\rm Diff}(M)$ and any two foliations are
strongly equivalent.  The gauge subgroup ${\rm Diff}_G(M,s)$ of
${\rm Diff}(M,s)$ depends on the specific theory under consideration.

The space $\Cb$ of quantum histories is now the moduli space of
`generalized connections' defined as follows \cite{8}. Let
${\A}_{\rm W}$ denote the $C^\star$ algebra generated by Wilson loop
functions (i.e. traces of holonomies of smooth connections around
closed loops in $M$). $\Cb$ is the Gel'fand spectrum of ${\A}_{\rm W}$.
Therefore, it is naturally endowed with the structure of a compact,
Hausdorff space and one can show that the moduli space of smooth
connections ${\cal C}$ is naturally and densely embedded in $\Cb$.

The label space $\cal L$ consists of triples
$f=(\gamma,\vec{j},\vec{I})$ where $\gamma$ is a graph in $M$,
$\vec{j}$ a labeling of its edges with non-trivial irreducible
equivalence classes of representations of $K$, and $\vec{I}$ a
labeling of its vertices with intertwiners.  The stochastic process is
defined by $f\mapsto P_f(\phi):=T_{\gamma,\vec{j},\vec{I}}(\phi)$
where the latter is a spin-network function on $\Cb$ \cite{5,jb}.
(Roughly, each $\phi\in\Cb$ assigns to every edge of $\gamma$ a group
element, the representations $\vec{j}$ convert these elements into
matrices and the function $T_{\gamma, \vec{j}, \vec{I}}$ arises from
contractions of indices of these matrices and intertwiners $\vec{I}$.
For details, see \cite{5,jb}.) Thanks to this judicious choice of
${\cal L}$, the measure $\mu_0$ can be defined quite simply:
\be 
<P_f>=0\,\, {\rm for\,\, all}\,\, f \,\,\,
{\rm except}\, f_0=(\emptyset,\vec{0},
\vec{0}),\quad {\rm and} \quad  <P_{f_0}>=1. \ee

It is easy to see that this measure satisfies axioms (II) and (III) :
The only property that one needs to use is that spin-network functions
form an orthogonal basis for ${\cal H}_{D+1}$. Given a foliation $E$,
the equivalence classes under the quotient by the null vectors are in
one to one correspondence with finite linear combinations of
spin-network states whose graph lies entirely in the surface
$E_0(\sigma)$. This then defines the Hilbert space ${\cal H}^E_D$
which is easily seen to be isomorphic to the Hilbert space defined by
the quantum configuration space \cite{8} over $\sigma$ and the
corresponding uniform measure $d\mu_{0,\sigma}$. 

While the uniform measure is associated with certain mathematical
models introduced by Husain and Kucha\v{r} \cite{hk}, it does not
capture the dynamics of a physical system. Therefore, we have some
freedom in the choice of ${\rm Diff}_G(M,s)$.  However, if we want to
satisfy both the remaining axioms, (GI) and (I), no choice is
entirely satisfactory. For example, every diffeomorphism in ${\rm
Diff}(M,s)$ has a non-trivial (unitary) action on ${\cal H}_D^E$ so
that axiom (GI) is satisfied only if we take the group of gauge
diffeomorphisms to be trivial.  For this choice of ${\rm
Diff}_G(M,s)$, and thus for any other, the measure $d\mu_0$ also
satisfies axiom (IV) \cite{9}.  However, while ${\rm Diff}(M,s)$ has
an unitary action on ${\cal H}_{D+1}$, there is no one-parameter group
of diffeomorphisms that acts {\it strongly continuously} on the
Hilbert space $L^2(d\mu_0)$ \cite{almmt3}. Thus, axiom (I) is
satisfied only for the complementary trivial choice ${\rm Diff}_G(M,s)
={\rm Diff}(M,s)$.  Nonetheless, it is true that $P_{f_0}=1$ is the
only state invariant under all time translations.

\subsection{Two-dimensional Yang-Mills Theory} 
\label{s4.2}

As mentioned in Section \ref{s2.2}, in two space-time dimensions,
Yang-Mills action requires only an area 2-form rather than a full
space-time metric. Therefore, the theory is invariant under all area
preserving diffeomorphisms and thus provides an interesting example
for our general framework.

Since connected, one-dimensional manifolds without boundary are
diffeomorphic either to the circle $S^1$ or to $\Rl$, let us consider
Yang-Mills theory with structure group $K = SU(N)$ on space-time
manifolds $M = \Rl \times \sigma$ where $\sigma = S^1$ or $\sigma
= \Rl$. The background structure $s$ is an area two-form $\omega$ on
$M$ and the action reads
\be
\la{B1}
S(A) = - \frac{1}{g^2} \int_M {\rm Tr}(F \wedge \star F) = 
- \frac{1}{g^2} \int_M {dx^0dx^1 \over \omega_{01}} tr(F_{01}^2) \ ,
\ee
where $F$ denotes the curvature two-form of the connection $A$ and
$(x^0,x^1) = (t,x)$ are the standard coordinates on $\Rl \times
\sigma$. In this case, the group $\difs$ is the group $\difo$ of area
preserving diffeomorphisms. The classical Hamiltonian formulation
shows that the gauge transformations of the theory correspond only to
local $SU(N)$-rotations. Thus, ${\rm Diff}_{G}(M,\omega)$ contains
only the identity diffeomorphism. Finally, the example given in
section IIB can be trivially adapted to the case under consideration
(simply by replacing $\Rl^3$ by $S^1$ and $\eta$ by $\omega$) to show
that there exist compatible foliations which fail to be strongly
equivalent. However, it is not difficult to show that all compatible
foliations are in fact weakly equivalent. We will now construct a
quantum field theory for this system, satisfying all our axioms.

For the standard area form, the reconstruction of the Hamiltonian
formalism from the Euclidean measure was obtained in \cite{almmt2}.
The particular Euclidean measure utilized was the limit as the lattice
spacing $a$ goes to zero of the Wilson lattice action for Yang-Mills
theory.  Let us recall some of the results adapted to the case of a
general area form.  As label space $\L$ we use $N$-1-tuples of loops
in $M$, $f = (\a_1, \cdots , \a_{N-1})$.  Let $\Cb$ be the moduli
space of generalized connections as in the previous example and the
random process be given by
\ba
\la{B2}
\L \ & \rightarrow & \ {\cal F}(\Cb)  \nonumber \\ 
f = (\a_1, \cdots , \a_{N-1}) \  & \mapsto & \ 
P_{(\a_1, \cdots , \a_{N-1})}(\phi) := T_{\a_1}(\phi) 
\cdots T_{\a_{N-1}}(\phi)
\ . \nonumber
\ea
Here $T_{\a}$ denotes the Wilson function,
$$
T_{\a}(\phi) = {1 \over N} {\rm Tr} (h_\a(\phi)) \ ,
$$
where $h_\a$ is the holonomy corresponding to the loop $\a$ 
and the (generalized) connection $\phi$. 

To begin with, let us consider any compatible foliation $E$. In
order to adapt the calculations of \cite{almmt2} we consider a
ultraviolet regulator $a$ by taking a (possibly) curved lattice in
$M$ made of plaquets diffeomorphic (as manifolds with boundary) to
rectangles, with area $a^2$ and such that the time-zero slice
$\gamma_E = E_0(\sigma)$ is a union of edges of plaquets.  Notice that
if $\sigma = S^1$, $\gamma_E$ is a (homotopically non-trivial) loop in
$M$.

It is easy to verify that the calculations and results of \cite{almmt2} remain
essentially the same and that we can take the ultraviolet limit $a
\rightarrow 0$ in the expression for the generating functional
$\chi(\a_1, \cdots , \a_{N-1}) = <T_{\a_1}(\phi) \cdots
T_{\a_{N-1}}(\phi)>$. Axioms II and III hold and so we can construct
the physical Hilbert spaces. Irrespective of the choice of the
compatible foliation $E$, the physical Hilbert space is
one-dimensional if $\sigma = \Rl$ and is $L^2(SU(N)/Ad_{SU(N)},
d\tilde \mu_H)$ if $\sigma = S^1$, where $\tilde \mu_H$ is the measure
induced on $SU(N)/Ad_{SU(N)}$ by the Haar measure on $SU(N)$.

Let us concentrate on the more interesting case of $\sigma = S^1$. From
\cite{almmt2} we obtain that the time evolution operator $\hat C_E^t$
is given by
\be
\la{B3}
[\hat C_E^t] = e^{{1 \over 2} g^2 {\rm Area}(E,t) \Delta} \ ,
\ee
where $\Delta $ denotes the invariant Laplacian on $SU(N)$ (functions
on \break $SU(N)/Ad_{SU(N)}$ can be thought as $Ad_{SU(N)}$-invariant
functions on $SU(N)$) and ${\rm Area}(E,t)$ denotes the area inclosed
between the loops $\gamma_E = E_0(S^1)$ and $E_t(S^1)$.  Thus:
\be
\la{B4}
{\rm Area}(E,t) = \int_0^t dt' \int_0^1 dx (E^* \omega)_{01}(x,t') \ .
\ee
Since $\vp^u_E = E \circ \vp^u \circ E^{-1} \in \difo \ , \forall u
\in \Rl$ (where $\vp^u$ denotes the standard time translation on
$\rs$) or, equivalently, $\vp^u \in {\rm Diff}(\Rl \times S^1 , E^*
\omega) \ , \ \forall u \in \Rl$, the component $(E^* \omega)_{01}$
does not depend on $t$. Therefore the area in (\ref{B4}) is linear
in $t$. Now, we showed in Lemma 5 that $[\hat C_E^t] = exp(- t [\hat
H^E])$. Hence, the Hamiltonian can now be read-off as:
\be
\la{B5}
[\hat H^E] = - {1 \over 2} g^2 L_E \Delta \ ,
\ee
where
\be
\la{B6}
L_E = \int_0^1 dx (E^* \omega)_{01}(x) \ .
\ee
Notice that if $\tilde E$ is strongly equivalent to $E$,
  $L_{\tilde E} = L_{E}$ and the two Hamiltonians 
agree as expected from Theorem 1. 

What would happen if we use a foliation $\tilde{E}$ which is {\it not} strongly
equivalent to $E$? Then, the value of $L_{E}$ (and therefore also
the Hamiltonian $[\hat H^{E}]$) will in general change.  For
example, if we choose, as in section IIB, $\tilde{E}: \Rl\times S^1
\rightarrow M, \, (t,x) \mapsto \tilde{E}_t(x) = (bt,x)$ with $b>0$,
then $L_{\tilde E} = b L_{E}$ and therefore $[\hat H^{\tilde E}] = b
[\hat H^{E}]$. This is, however, exactly what one would expect since
the vector field generating the time translation on $M$ defined by $E$
is $b$ times that defined by $E$. This is a concrete illustration of
remark b) at the end of sec \ref{s3}.

Finally, in this model, the axiom (GI) holds trivially since ${\rm
Diff}_{G}(M, \omega)$ contains only the identity
diffeomorphism. Furthermore, all physical states are manifestly
$SU(N)$-gauge invariant. The existence of a Hamiltonian operator
$[\hat{H}^E]$ implies that axiom (I) holds. Axiom (IV) also holds and
the vacuum state is unique.

\subsection{2+1 gravity and BF-Theories}
\label{s4.3}

Fix a 3-manifold $M$ with topology $\Rl \times \sigma$, where $\sigma$
is a compact 2-manifold. In the first order form, the basic fields for
general relativity can be taken to be a connection $A$ and a
Lie-algebra-valued 1-form $e$. The action is given by
\be \label{2+1}
S(A,e) = \int_{M}\, {\rm Tr}\,\, e\wedge F
\ee
where the trace is taken in the fundamental representation. If the
structure group $K$ is $SO(3)$, we obtain general relativity with
signature +,+,+ while if the structure group is $SO(2,1)$, we obtain
general relativity with signature -,+,+. The field $e$ can be thought
of as a triad, and when $e$ satisfies the equation of motion, the
field $A$ is the spin-connection compatible with the triad. The
equations of motion on $A$ say that $F$ vanishes. In this case, there
is no background structure $s$, ${\rm Diff}_G(M)$ is the connected component
of the identity of ${\rm Diff}(M)$, and all foliations are strongly
equivalent.

The heuristic measure on the space of paths $(A,e)$ is given by
`${\cal D}\!A\, {\cal D}\!e\, \exp iS(A,e)$' and if we integrate out the
$e$ fields we obtain the measure `$\delta(F)\, {\cal D}\!A$' on the
space of connections. This suggests that the rigorous measure should
be concentrated on flat connections. It turns out that the moduli
space of flat connections is a finite dimensional symplectic manifold%
\footnote{There are certain technical subtleties in the $SO(2,1)$ 
case \cite{10}. In what follows we will assume that a Hausdorff
manifold has been obtained by deleting suitable points. The resulting
moduli space has disconnected components. By moduli space we will
refer either to the `time-like' or `space-like' components.}
and therefore has a natural Liouville measure. 

With this intuitive picture in mind, we will now construct a quantum
field theory for this system, satisfying all our axioms.  Choose for
$\Cb$ the moduli space of smooth connections (or a suitable completion
thereof. For example, in the $SO(3)$ theory one can use the completion
used in example 1.) For ${\cal L}$ we use the space of closed loops
$f$ on $M$. The stochastic process is defined by $f\mapsto P_f(\phi) =
{\rm Tr}\, h_f(\phi)$ where $h_f(\phi)$ is the holonomy of $\phi \in
\Cb$ around the closed loop $f$ in $M$ and trace is taken in the
fundamental representation. The measure is defined by
\be <\chi_f>\, =\,  <P_f(\phi)> \, =\,  \int_{{\cal M}_o} 
d\mu_L P_f(\phi), 
\ee
where ${\cal M}_o$ is the moduli space of flat connections and $\mu_L$
is the Liouville measure thereon. Note incidentally that, in the
resulting history Hilbert space ${\cal H}_{D+1}$, $P_f$ and $P_{f'}$
define the same element if $f$ and $f'$ are homotopic to each other.
Hence ${\rm Diff}_G(M)$ is represented by the identity operator on ${\cal
H}_{D+1}$.

It is straightforward to check that the axioms (II), (III), (GI), (I)
and (IV) are all satisfied. (In fact (I) and (IV) hold trivially
because ${\rm Diff}_{G}(M,s)$ is so large.)  The Hilbert space
${\cal H}_D^E$ is isomorphic to $L^2({\cal M}_o, d\mu_L)$. The
Hamiltonian theory can be constructed independently through canonical
quantization \cite{11} and yields precisely the same Hilbert space of
physical states. Note that the correct correspondence between the path
integral and canonical quantization holds for both signatures,\, -,+,+
and +,+,+. However, one has to use measures whose heuristic analogs
involve $\exp iS$ in {\it both} cases, so that the signature of the
associated metric is not fundamental to determining the heuristic form
of the measure.  In particular, the Wick rotation has no obvious role
in the diffeomorphism invariant context. (For further discussion, see
\cite{2}.)

This viewpoint is also supported by the fact that 2+1 dimensional
general relativity is a special case of B-F theories which can be
defined in any dimension and in which there is no natural metric at
all; the presence of a metric can thus be regarded as an `accident'
of 2+1 dimensions. In these theories, the basic fields are a
connection $A$ and a D-1 form $B$ with values in the dual of the Lie
algebra. The action has the same form as (\ref{2+1}) with $e$ replaced
by $B$. One can repeat essentially the same construction for all of
these theories.

\section{Discussion}
\label{s5}

In this paper, we introduced an extension of the Osterwalder-Schrader
framework to diffeomorphism invariant theories.  The key idea was to
generalize the standard setting by dropping all references to the
space-time metric. We considered $D+1$ dimensional space-times $M$
with topology $\Rl\times \sigma$, where $\sigma$ is allowed to be an
arbitrary, $D$-dimensional manifold. Heuristically, $\Rl$ serves as a
generalized `time direction'. More precisely, using foliations $E$ of
$M$, with leaves transverse to the $\Rl$-direction, we were able to
extend the standard notions of time translation and time reflection
without any mention of a space-time metric. This in turn enabled us to
generalize the Osterwalder-Schrader axioms and construct a Hamiltonian
quantum theory starting from a path integral.  While $M$ is required
to have a product topology, given our goal of constructing a
Hamiltonian framework, this restriction is unavoidable.

As in the original Osterwalder-Schrader framework, the key
mathematical object in the path integral formulation is the measure
$\mu$ on the space $\Cb$ of quantum histories and the axioms are
restrictions on permissible $\mu$. In the construction of a bridge
from the path integral to the Hilbert space theory, two of these axioms
play a central role: reflection positivity (axiom III, unchanged from
the original Osterwalder-Schrader treatment) and diffeomorphism
invariance (axiom II, which replaces the Euclidean invariance of the
standard treatment). Given a foliation $E$ of $M$ and a measure $\mu$
satisfying reflection positivity, one can construct the Hilbert space
${\cal H}^E_D$ of quantum states. The diffeomorphism invariance of
$\mu$ then ensures that the Hilbert space is essentially insensitive
to the choice of the foliation $E$. The remaining axioms ensure the
existence of the Hamiltonian operators generating (generalized)
time-translations which are true (i.e. non-gauge) symmetries of the
theory and the existence and uniqueness of a vacuum state.

Perhaps the most striking feature of the present framework is its
generality.  We did not have to restrict ourselves to specific
space-time manifolds and the Lagrangian --- indeed even the matter
content--- of the theory was left arbitrary. In particular, our
generalized setting allows theories of interacting gauge and tensor
fields with arbitrary index structure, general relativity, higher
derivative gravity theories, etc. However, this generality comes at a
price.  As with the original Osterwalder-Schrader reconstruction
theorem, the results of this paper only tell us how to obtain a
Hilbert space theory from a given measure satisfying certain axioms.
It does \textit{not} tell us how to construct this measure from a
given classical theory.  For familiar field theories (without
diffeomorphism invariance), $\exp -\,S_E$, with $S_E$, the Euclidean
action, generally provides a heuristic guide in the construction of
this measure. We saw in Sections \ref{s4.2} and \ref{s4.3} that, in
the absence of a space-time metric, the distinction between the usual
Euclidean and Lorentzian prescriptions become blurred. In some cases,
the heuristic guide is again provided by $\exp -S$ while in other
cases it is provided by $\exp iS$. Thus, the construction of the
measure now acquires a new subtle dimension. Furthermore, because our
setting is much more general than the original one, even the
`kinematical structure' ---the spaces $\Cb$ of quantum histories,
the label set ${\cal L}$ and pairings $P$ of Section \ref{s2}--- can
vary from one theory to another and have to be constructed case by
case. However, for diffeomorphism theories of connections, including
general relativity in three space-time dimensions, we were able to
provide natural candidates for these structures and find the
appropriate measures. In \cite{2}, we will extend these considerations
to more general contexts, albeit at a more heuristic level. In
particular, starting from the classical Hamiltonian framework, we will
discuss how one can construct heuristic measures. We will find some
subtle but important differences from the familiar cases.

\bigskip\bigskip {\bf Acknowledgments:} We thank Jerzy Lewandowski for
stimulating discussions, Olaf Dreyer for comments on an early draft
and Jorge Pullin for numerous suggestions which significantly improved
the final presentation. This research project was supported in part
by the National Science Foundation under grant PHY95-14240 to Penn
State, PHY97-22362 to Syracuse University and PHY94-07194 to ITP,
Santa Barbara, by the Eberly research funds of Penn State, by research
funds from Syracuse University, by CENTRA/IST, by projects
PRAXIS/2/2.1/FIS/286/94 and CERN/P/FIS/1203/98.

\begin{appendix}

\section{A Universality Class of Generally Covariant Quantum Gauge Field
Theories in Two Spacetime Dimensions}
\label{sa}

The purpose of this appendix is to show that, if $D=2$, the uniform
measure $\mu_0$ on the space $\overline{\cal C}$ of generalized
connections considered in section \ref{s4.1} naturally arises in the
path integral formulation of a class of diffeomorphism invariant gauge
theories. For generalizations and further discussion, see \cite{12}.

Let $M$ be a two-dimensional manifold with topology $\Rl^2$ or
$S^1\times \Rl$, and let $G$ be a compact, connected, semi-simple
gauge group. We will denote by $A$ the pull-back by local sections
of a connection on a principal $G$-bundle over $M$ and by $F$ its
curvature. We will use $a,b,c,..=1,2$ as the tensorial indices and
$i,j,k,..=1,..,\dim(G)$ as the Lie algebra indices. Choose a basis
$\tau_i$ of the Lie algebra $Lie(G)$ of $G$ normalized such that
tr$(\tau_i\tau_j)=-N\delta_{ij}$ and structure constants are defined
by $[\tau_i,\tau_j]=2f_{ij}\;^k \tau_k$.  Finally, let
$\epsilon^{ab}$ be the metric independent totally skew tensor density
of weight one and use it to represent the curvature as a Lie-algebra
valued scalar density $F^i:=\frac{1}{2}\epsilon^{ab} F_{ab}^i$.

Let us consider the following action 
\be \label{a.1} S=
\int_M \,d^2x \,\left[F^i F^i\right]^{1\over 2} 
\equiv \int_M \, d^2x  \left[P_2(F)\right]^{1\over 2} 
\ee 
where $F^iF^i=k_{ij}F^iF^j$ is the norm of $F^i$ with respect to the
Cartan-Killing metric.  Since the quantity under the square-root is a
scalar density of weight two, the action (\ref{a.1}) is diffeomorphism
invariant. This is perhaps the simplest of such actions for
$G$-connections. Other, more general diffeomorphism invariant actions
can be constructed by taking n-th roots of suitable n-nomials in $F^i$
and their covariant derivatives.

Assuming appropriate boundary conditions, the equations of motion
that follow from (\ref{a.1}) are 
\be \label{a.2}
D_a\frac{F^i}{\sqrt{F^j F^j}}=0
\ee
These equations are consistent since the integrability condition
$\epsilon^{ab} D_a D_b (F^i/\sqrt{F^j F^j})= f^i\;_{jk} F^j
F^k/\sqrt{F^l F^l}=0$ is identically satisfied.  The general solution
is $F^i=s b^i$ where $s$ is an arbitrary, nowhere negative, scalar
density of weight one and $b^i$ a covariantly constant vector of unit
norm. In the special case $G=U(1)$, we have: $b^i(x)=b^1(x)=\pm 1$ must be
constant, say $b^1(x)=1$ and thus $F(x)=\epsilon^{ab} (\partial_a
A_b)(x) \ge 0$ is the general solution. The case $G=U(1)$ also
gives a simple class of solutions for general $G$ : Suppose we
choose the connection to be of the form $A_a^i=a_a t^i$ where $t^i$ is
a constant unit norm vector. Then $F^i=F t^i$ where $F=\epsilon^{ab}
\partial_a a_b$ and $D_a F^i/{\sqrt{F^j F^j}}=t^i \partial_a F/|F|$
and this reduces the problem to $G=U(1)$. Thus, the space of
solutions to the field equations is infinite dimensional.

We now wish to derive a path integral for this theory, i.e., construct
a continuum measure along the constructive approach of \cite{almmt2},
using the action (\ref{a.1}) as the classical input. Let us focus on
the non-trivial case when $M$ is topologically $\Rl\times S^1$.  Let
us introduce a system of coordinates $(x,t)$ where $x\in[-a,a]$
denotes the compact direction and  $a$ is an arbitrary parameter of
dimension of length. Next we introduce a foliation of the cylinder by
circles of constant `time' $t$ coordinate. We introduce a
dimensionless UV cut-off $\epsilon$ and an IR cut-off $T$ of dimension
of length.  Consider $1+2N,N=T/(\epsilon a)$ circles at values of $t$
given by $t/a=l\epsilon,\;l=-N,-N+1,..,N$ and similarly
$2N'+1,N'=1/\epsilon$ coordinate lines at constant values of $x$ given
by $x/a=k\epsilon,\;k=-N',-N'+1,.., N'$ with $x/a=\pm 1$
identified. Thus we have covered a portion $M_T$ of the cylinder $M$,
corresponding to $(x,t)\in [-a,a]\times [-T,T]$ by a lattice
$\gamma_{\epsilon,T}$ of cubic topology and can consider the usual
plaquette loops $\Box$ of that lattice based at the point $p$ with
coordinates $x=t=0$, say. We can label plaquettes by two integers
$(k,l)$ in the obvious way and have \be \label{a.3}
\Box_{k,l}=\rho_{k,l}\circ e_{k,l}\circ f_{k+1,l}\circ e_{k,l+1}^{-1}
\circ f_{k,l}^{-1}\circ\rho_{k,l}^{-1} \ee where $e_{k,l},\;f_{k,l}$
are respectively edges of the lattice at constant values of $t$ and
$x$ respectively and $\rho_{k,l}$ is an arbitrary but fixed lattice
path between the base point and the corner of the box corresponding to
lowest values of $x,t$ respectively (modulo the identification of
$x/a=\pm 1$).  If we now parameterize edges as the image of the
interval $[0, a]$ then it is not difficult to see that for a
classical connection $A$ the holonomy for a plaquette is to leading
order in $\epsilon$ given by 
\be \label{a.4}
h_{\Box_{k,l}}=1+a^2\epsilon^2 \epsilon_{ab}\dot{e}_{k,l}^a(0)
\dot{f}_{k,l}^b(0) F^j(x=k\epsilon a,t=l\epsilon a)\tau_j/2 \ee 
Then,
\be \label{a.5} S_{\epsilon,T}:= \sum_{k,l} \left[ P_2(-\frac{2}{N}
\mbox{tr}(\tau_i(h_{\Box_{k,l}}-1)))\right]^{1\over 2}=: \sum_{k,l}
s(h_{\Box_{k,l}}) \ee
can be taken to be a Wilson-like action that approximates (\ref{a.1})
in the sense that 
\be \label{a.6} \lim_{\epsilon\to 0}
S_{T,\epsilon}= \int_{M_T}\, d^2x\, \left[{P_2}(F)\right]^{1\over 2} 
\ee
The elementary but important observation is that function $s(h)$
defined in (\ref{a.5}) is the same for all plaquette loops.

Let us choose as our random variables the `loop network functions'
$T_{\vec{\alpha},\vec{\pi},c}$ \cite{13}.  These are similar to the
$T_{\gamma, \vec{j}. \vec{I}}$ considered in section \ref{s4.1}),
except that: i)now the graphs $\alpha$ are replaced by a finite
collection, say $L$, of mutually non-overlapping loops $\vec{\alpha}$
based at $p$ (possibly including the homotopically non-trivial one
that wraps once around the cylinder at $t=0$); ii) $\vec{\pi}$ now
denotes a collection of $L$ equivalence classes of non-trivial
irreducible representations of $G$ subject to the constraint that if
loops $\alpha_{I_1},..,\alpha_{I_r},r\le L,I_1<..<I_r$ share a
segment, then the tensor product $\pi_{I_1}\otimes...\otimes \pi_{I_r}$
does not contain a trivial representation; and, iii) $c$ is an
intertwiner between the trivial representation and
$\pi_1\otimes..\otimes\pi_L$. The function
$T_{\vec{\alpha},\vec{\pi},c}(A)$ depends on $A$ through the
holonomies $h_{\alpha_I}(A),\;I=1,..,L$ only. Given any measure on the
moduli space $\overline{\cal C}$ of generalized connections, one can
compute the the expectation values of these loop-network
functions. These provide us with the characteristic function of the
underlying measure \cite{7,almmt2}.

Using this machinery, we can write down the regularized measure on
$\overline{\cal C}$ by specifying its characteristic function:
\be \label{a.7}
<T_{\vec{\alpha},\vec{\pi},c}>_{T,\epsilon} 
:=\frac{1}{Z_{\epsilon,T}}
\prod_{k,l} ([\int_G d\mu_H(h_{e_{k,l}})]\;[\int_G d\mu_H(h_{e_{k,l}})])
e^{-S_{\epsilon,T}} T_{\vec{\alpha},\vec{\pi},c}\, , \ee
where,
\be Z_{\epsilon,T}:= \prod_{k,l} ([\int_G
d\mu_H(h_{e_{k,l}})]\;[\int_G d\mu_H(h_{e_{k,l}})])
e^{-S_{\epsilon,T}}\, .  \ee
Here, of course, all the loops in question live on our lattice.

In order to explicitly compute the expectation value (\ref{a.7}) in
the limit $\epsilon\to 0$ and $T\to\infty$ we can essentially follow
\cite{almmt2}. This is due to a peculiarity of $D=1$ and the planar or
cylindrical topology of $M$ namely, that the plaquette loops are
holonomically independent. Let then, at fixed $\epsilon,T$, $|\alpha|$
be the number of plaquettes contained in the surface bounded by
$\alpha$. Then, repeating literally all the calculations performed in
\cite{almmt2} we find 
\ba \label{a.8}
<T_{\vec{\alpha},\vec{\pi},c}>_{T,\epsilon} &=&
\,\,T_{\vec{\alpha},\vec{\pi},c}(A=0) \prod_{I=1}^L
(\frac{J_{\pi_I}}{J_{\pi_0}})^{|\alpha_I|}\nonumber\\ J_\pi &=&
\frac{1}{\dim(\pi)}\int_G d\mu_H(h) \chi_\pi(h) e^{-s(h)} \ea 
Here $\chi_\pi$ denotes the character of $\pi$,
$\chi_\pi(1)=\dim(\pi)$ and $\pi_0$ denotes the equivalence class of
the trivial representation.  If one of the loops, say $\alpha_I$, is
homotopically non-trivial then
$(\frac{J_{\pi_I}}{J_{\pi_0}})^{|\alpha_I|}$ has to be replaced by
$\delta_{\pi_I,\pi_0}$.

The UV and IR cut-off can now be trivially removed from (\ref{a.8}) :
Due to the holonomic independence of the plaquette loops the quantity
(\ref{a.8}) is already independent of $T$.  Due to the diffeomorphism
invariance of the original action, the quantity (\ref{a.8}) depends on
$\epsilon$ only through the numbers $|\alpha_I|$ which just counts the
number of plaquette loops that one uses in order to approximate the
loop $\alpha_I$. In contrast to the situation with two-dimensional
Yang-Mills theory \cite{almmt2}, the numbers $J_\pi$ are already
independent of $\epsilon$. The reason for this is, of course, the
background independence of the original action (\ref{a.1}). By
contrast, as we saw in section \ref{s4.2}, the Yang-Mills action
requires a background area-element. At first the difference seems
small. However, it leads one in Yang-Mills theory to the Wilson action
$\frac{1}{\epsilon^2} \sum_\Box [s(h_\Box)]^2$ ---rather than $\sum_\Box
s(h_\Box)$--- which makes $J_\pi$ $\epsilon$-dependent in just the
right way to produce the area law \cite{almmt2} as $\epsilon\to 0$.
In the present case there is no background structure and therefore we
cannot have an area law; there is no area 2-form to measure the area
with! Instead we have the following. Since the definition of
$J_\pi$ implies $|J_\pi|<J_{\pi_0}$ for $\pi\not=\pi_0$ and since in
the limit $\epsilon\to 0$ the numbers $|\alpha_I|$ diverge, it follows
that $\lim_{\epsilon\to 0}
(J_{\pi_I}/J_{\pi_0})^{|\alpha_I|}=\delta_{\pi,\pi_0}$. Consequently,
we have:
\be \label{a.9}
\lim_{T\to\infty} \lim_{\epsilon\to 0} 
<T_{\vec{\alpha},\vec{\pi},c}>_{T,\epsilon} 
=\left\{  \begin{array}{cc}
1 & \mbox{ if } \vec{\alpha}=\{p\},\vec{\pi}=c=\pi_0 \\
0 & \mbox{ otherwise}
\end{array} \right.
\ee
in other words, we arrive at the characteristic functional of the
uniform measure in $D=1$ discussed in section \ref{s4.1}. Note that
the limit (\ref{a.9}) is {\it completely insensitive to the choice of
the regularizing lattice}. Hence, the result is independent of the
regulator.

Finally, we note that in place of (\ref{a.1}) we could have considered
the most general diffeomorphism and gauge invariant action
$\tilde{S}(A)$ which depends on the field strengths $F^i$ but not on
their derivatives. For this, we can begin with globally defined
(i.e. gauge invariant) monomials
$F^n:=F^{i_1}..F^{i_n}\mbox{tr}(\tau_{i_1}..\tau_{j_1})$. (If the rank
of $G$ is $R$, only the first $R-1$ of these will be independent,
others being polynomials in them.) Let $P_n$ be an arbitrary, gauge
invariant, positive semi-definite, homogeneous polynomial of degree $n$
in the $F^j$ and set
\be \label{a.10}
\tilde{S}(A)=\sum_{n=1}^\infty a_n \int_M d^2x \sqrt[n]{P_n[F]}
\ee
where $a_i$ are non-negative constants all but a finite number of
which are zero. This action is diffeomorphism invariant, again because
the integrand is a scalar density of weight one.  We could have
carried out the above procedure for any of these
theories. Irrespective of the action in this class, the final
characteristic function would have been again (\ref{a.9}). Thus, all
these theories lie in the same universality class; their
renormalization group flows all reach the same UV fixed point and the
final quantum theory is dictated by the uniform measure. This issue, 
the Hamiltonian formulation, and the canonical quantization of these
theories will be discussed in \cite{12}.

\end{appendix}

\end{document}